\documentstyle[12pt,epsf]{article}
\if@twoside  \oddsidemargin 26pt \evensidemargin -18pt 
\medskipamount\newcommand{\be}{\begin{eqnarray}}
\begin{document}
\hbox{}
\nopagebreak\vspace{-3cm}
\addtolength{\baselineskip}{.8mm}\baselineskip=24pt

\begin{flushright}{\sc  TPI--MINN--97--20/T} \\{\sc    NUC--MINN--97--6/T} \\
{\sc    HEP--MINN--97--1546} \\
{\sf Cavendish-HEP-97/09}
\end{flushright}
\begin{center}{\Large \bf The Wilson renormalization group for low x
physics: towards the high density regime}\\
\vspace{0.1in}\vspace{0.5in}
{\large  Jamal Jalilian-Marian$^1$, Alex Kovner$^{1,2 *}$,
Andrei Leonidov$^3$, Heribert Weigert$^4$}\\
$^1$
{\it Physics Department, University of Minnesota, Minneapolis, MN 55455, USA}\\
$^2$
{\it Theoretical Physics, Oxford University, 
1 Keble Road, Oxford OX1 3NP, UK}\\
$^3$
{\it Theoretical Physics Department, P.N.~Lebedev Physics Institute,
\\Leninsky pr. 53  Moscow, Russia}\\
$^4$
{\it University of Cambridge, Cavendish Laboratory, HEP, 
Madingley Road,CB3 0HE UK}\\
\vspace{0.1in}\vspace{1in}
{\bf  Abstract} 
\end{center}

\noindent
We continue the study of the effective action for low $x$ physics
based on a Wilson renormalization group approach. We express the full
nonlinear renormalization group 
equation in terms of the average value and the average
fluctuation of extra color charge density generated by integrating out
gluons with intermediate values of $x$. This form clearly exhibits the
nature of the phenomena driving the evolution and should serve as the
basis of the analysis of saturation effects at high gluon density
at small $x$.

$^*$ On leave from PPARC Advanced Fellowship.
\vfill

\newpage

Understanding of hadronic systems at large energies is one of the most
exciting open problems in QCD today (see, e.g., a recent review
\cite{Le}).  The recent wave of interest was sparked by experimental
data on deep inelastic scattering \cite{HERA} showing a steep rise in
the cross section at small values of Bjorken $x$.  Theoretically it is
believed that this rise is mainly triggered by fast growth of gluon
density in a proton and indeed most fits to the HERA data use gluonic
distributions that grow as a power at small $x$ \cite{MRS}.  It 
argued a long time ago \cite{GLR} that eventually the
rate of growth of the gluon density should slow down and eventually
saturate, thus curing a potential conflict with unitarity of the
underlying scattering. Physically this should happen as the
gluonic system becomes dense and gluons start overlapping in space causing
self interactions, which are normally neglected in a linear QCD
evolution, to become important.  The most exciting problem of low $x$
physics therefore is the question of how to deal with finite
density partonic systems.

A formulation which is naturally directed towards this type of
problems was suggested by McLerran and Venugopalan \cite{larry} in the
context of large nuclei.  Their approach was considerably modified and
generalized later \cite{JKMW}, \cite{linearRG} and evolved into an
effective action approach to the low $x$ hadronic structure.  In this
paper we continue to study this effective action with the help of
Wilson renormalization group.  The general Wilsonian renormalization
group procedure for study of the low x physics was discussed first in
\cite{JKMW}, and then in more detail in \cite{linearRG}. It was also
shown in the latter paper that in the linear regime, i.e.  for small
partonic densities, this renormalization group equation reduces to the
celebrated BFKL equation for the evolution of unintegrated gluon
density \cite{BFKL}.  This is a very important test for the
consistency of the whole approach, however our goal is to be able to
deal with the nonlinear regime where the color charge densities are
not small and therefore have to be treated nonperturbatively.

In this note we take one further step in this direction. The
renormalization group procedure involves functional integration over
two sets of variables: the fluctuation modes of the gluon field with
longitudinal momenta above the cutoff and the surface color charge
density $\rho$ at fixed value of $\rho^\prime$ which then becomes the
charge density in the effective theory that describes only low
longitudinal momentum glue modes. The goal of this note is to show how
to perform one part of this program, namely the integration over the
color charge density distribution. This will result in a gratifyingly
compact and simple repesentation of the full nonlinear RG equation in
terms of the average value and the average fluctuation of extra color
charge density generated by integrating out gluons with intermediate
values of $x$. The evolution towards smaller $x$ is understood therefore in
a simple way in terms of these two physical quantities. 
The simple form of the
result is ideally suited to discuss the validity and the limitations of the
BFKL equation from a new vantage point, as the low densitiy limit of
the full nolinear RG equations.

First, let us recall the framework as set up in \cite{linearRG}.
The starting point is the following action given in
the Light Cone gauge $A^+ = 0$ 
\begin{eqnarray}
S&=&i\int d^2 x_t F[\rho ^a(x_t)]\\
&-& \int d^4 x {1\over 4}G^2 + {{i}\over{N_c}} \int d^2 x_t dx^-
\delta (x^-)
\rho^{a}(x_t) {\rm tr}T_a W_{-\infty,\infty} [A^-](x^-,x_t)\nonumber 
\label{action}
\end{eqnarray}

Here $G^{\mu \nu} $ is the gluon field
strength tensor
\begin{equation}
G^{\mu \nu}_{a} = \partial^{\mu} A^{\nu}_{a} 
- \partial^{\nu} A^{\mu}_{a} + 
g f_{abc} A^{\mu}_{b} A^{\nu}_{c}\nonumber
\end{equation}
$T_a$ are the $SU(N)$ color matrices in the adjoint representation
and $W$ is the path ordered exponential along the $x^+$ direction in
the adjoint representation of the $SU(N_c)$ group
\begin{equation}
W_{-\infty,\infty}[A^-](x^-,x_t) = P\exp \bigg[-ig \int dx^+
A^-_a(x^-,x_t)T_a \bigg] \end{equation}

The average of any gluonic operator $O(A)$ in the hadron is
calculated as 
\begin{equation} <O>={\int D[\rho^a,A^\mu_a]O(A)\
\exp\{iS[\rho, A]\}\over \int D[\rho^a,A^\mu_a] 
\exp\{iS[\rho, A]\}}
\label{average}
\end{equation}
The exponential of the imaginary part of the action
\begin{equation}
{\rm Im}\ S=\int d^2x_t F[\rho^a]
\end{equation}
can be thought of as a kind of ``free energy'' for the ensemble of the
color charge density $\rho(x_t)$.
The ``Boltzmann factor'' 
\begin{equation}
\exp\left(-\int d^2x_t F[\rho^a]\right)
\label{boltzmann}
\end{equation}
appearing in Eq.(\ref{average}) controls the statistical weight of a
particular configuration of the two dimensional color charge density
$\rho^a(x_t)$ inside the hadron.

In the classical approximation to the path integral in 
Eq.(\ref{average})\cite{larry} one finds 
the classical solution for the equations of motion 
at fixed $\rho$ , and then
averages over the charge density distribution with the ``Boltzmann weight''
Eq.(\ref{boltzmann}). 
The classical solution for any fixed $\rho(x_t)$ has the structure\footnote
{To be more precise, there is an infinite number of solutions to the
classical equations of motion at fixed $\rho$. This is a consequence of the
residual gauge invariance of the action. In the following we will be
working in the gauge $\partial_iA_i(x^-\rightarrow-\infty)=0$. This is a 
complete gauge fixing and it therefore picks a unique solution of the equations
of motion.}
\begin{eqnarray}
A^-_{cl}&=&0\nonumber \\
A^i_{cl}\equiv b^i&=&\theta(x^-)\alpha^i(x_t)
\end{eqnarray}
where $\alpha_i$ is related to $\rho$ through
\begin{eqnarray}
\alpha_i(x_t)&=&{i\over g}U(x_t)\partial_iU^\dagger (x_t) \nonumber\\
\partial_i\alpha_i&=&- g\rho
\end{eqnarray}
and $U(x_t)$ is a unitary matrix.

The quantum corrections to this classical approximation are large at small
longitudinal momenta \cite{AJMV}. To resum these large corrections we follow 
the renormalization group
procedure as developed in \cite{JKMW},\cite{linearRG}.

Let us introduce the following decomposition of the gauge field:
\begin{equation}
A^a_{\mu} (x) = b^a_{\mu} (x) + \delta A^a_{\mu} (x) + a^a_{\mu} (x)
\end{equation}
where $ b^a_{\mu} (x)$ is the solution of the classical equations of
motion, $ \delta A^a_{\mu} (x) $ is the fluctuation field containing
longitudinal momentum modes $q^+$ such that $P^+_n < q^+ < P^+_{n-1}$
while $a_\mu$ is a soft field with momenta $k^+<P^+_n$, with respect
to which the effective action is computed. The initial path integral
is formulated with the longitudinal momentum cutoff on the field
$\delta A$, $q^+<P^+_{n-1}$.  The effective action for $a^\mu$ is
calculated by integrating over the fluctuations $\delta A$. This
integration is performed with the assumption that the fluctuations
are small as compared to the classical fields $b^a_{\mu}$. More
quantitatively, at each step of this RG procedure the scale $P^+_n$ is
chosen such that $\ln {P^+_{n-1}\over P^+_n}>1$, but $\alpha_s\ln
{P^+_{n-1}\over P^+_n} \ll 1$.  Expanding the action around the
classical solution $b^a_{\mu} (x)$ and keeping terms of the first and
second order in $\delta A$ we get
\begin{equation}
S= -{1\over 4}G(a)^2-{{1}\over{2}} 
\delta A_{\mu} [{\rm D}^{-1} (\rho)]^{\mu\nu} \delta A_{\nu} + 
ga^- \rho^\prime+O((a^-)^2)
+ iF[\rho]
\label{effectiveaction}
\end{equation}
where
\begin{equation}
\rho^\prime =\rho+ \delta \rho_1 + \delta \rho_2
\label{prime}
\end{equation}
with
\begin{eqnarray}
\delta \rho^a_1(x_t,x^+) &=& -2 f^{abc} \alpha^{b}_{i}\delta A^{c}_{i}(x^-=0)
\label{rho1}
\\ 
&-& {{g}\over{2}} f^{abc} \rho^{b}
(x_t) \int dy^+ \Bigg[\theta (y^+ - x^+) - \theta (x^+ - y^+) \Bigg] 
\delta A^{-c}(y^+,x_t, x^-=0)\nonumber
\end{eqnarray}
and
\begin{eqnarray}
\delta \rho^a_2(x) &=& -f^{abc} [\partial^+ \delta A^{b}_{i}(x) ]\delta A^{c}_{i}(x) \nonumber\\
&-& 
{{g^2}\over{N_c}} \rho^{b}(x_t) \int dy^+ \delta A^{-c}(y^+,x_t,x^-=0) \int dz^+ \delta A^{-d}(z^+,x_t,x^-=0) \nonumber \\
&\times &\Bigg[\theta (z^+ -y^+) \theta (y^+ -x^+)  {\rm tr} T^a T^c T^d T^b  \nonumber \\
&+&\theta (x^+ -z^+) \theta (z^+ -y^+)  {\rm tr} T^a T^b T^c T^d  \nonumber \\
&+&\theta (z^+ -x^+) \theta (x^+ -y^+)  {\rm tr} T^a T^d T^b T^c \Bigg]
\label{rho2}
\end{eqnarray}
The first term in both $\delta\rho_1^a$ and 
$\delta \rho^a_2$ is coming from expansion of $G^2$ in 
the action
while the rest of the terms proportional to $\rho$
are from the expansion of the 
Wilson line term. 
The three terms
correspond to different time ordering of the fields. 
Since the longitudinal momentum of $a^-$ is much lower than of $\delta A$, we 
have only kept the eikonal coupling (the coupling to $a^-$ 
component of the soft field), which gives
the leading contribution in this kinematics.
The inverse propagator $[{\rm D}^{-1}]^{\mu\nu}$ is given by
\begin{eqnarray}
\left[{\rm D}^{-1}\right]^{ij}_{ab}(x,y)
&=&\left[D^2(b)\delta^{ij}+D^i(b)D^j(b)\right]_{ab}\delta^{(4)}(x,y)
\\
\left[{\rm D}^{-1}\right]^{i+}_{ab}(x,y)
&=&-[\partial^+_xD^i_{ab}(b)\delta^{(4)}(x,y)+2f_{abc}\alpha_c^i(x_t)
\delta(x^-)\delta(y^-)\delta^{(2)}(x_t,y_t)\delta(x^+,y^+) ]\nonumber\\
\left[{\rm D}^{-1}\right]^{++}_{ab}(x,y)&=&(\partial^+)^2\delta_{ab}\delta^{(4)}(x,y)+
f_{abc}\rho^c(x_t)\delta(x^-)\delta(y^-)\theta(x^+-y^+)\delta^{(2)}(x_t,y_t)
\nonumber
\end{eqnarray}

The procedure now is the following:
\begin{enumerate}
\item Integrate over $\delta A^\mu$ at fixed $\rho$ and fixed
  $\delta\rho$.
\item  Integrate over $\rho$ at fixed $\rho^\prime=\rho+\delta\rho$.
\end{enumerate}
This generates the new effective action which formally can be written as
\begin{equation}
\exp\left(iS[\rho^\prime,
  a^\mu]\right)=\exp\left(-F^\prime[\rho^\prime]-{i\over 4}G^2(a)
  +iga\rho^\prime\right) 
\end{equation}
with
\begin{equation}
\exp\left(-F^\prime[\rho^\prime]\right)=
\int D[\rho,\delta A]\  \delta (\rho^\prime-\rho-\delta \rho[\delta A])
 \exp\left(-F[\rho]
-{i\over 2} \delta A D^{-1} [\rho] \delta A\right)
\label{newf}
\end{equation}
Of course, to leading order in $\ln 1/x$ only terms linear in $\alpha_s\ln 1/x$
should be kept in $F^\prime$. Defining
\begin{equation}
\label{Delta}\alpha_s\ln {1\over x}\Delta[\rho]\equiv F^\prime[\rho]-F[\rho]
\end{equation}
gives the RG equation
\begin{equation}
{d\over dy}F[\rho]=\alpha \Delta[\rho]
\label{evol}
\end{equation}

In order for Eq.(\ref{evol}) to be a bona fide renormalization group 
equation, we should be
able to find the functional $\Delta$ for arbitrary Boltzmann weight $F$.
That requires being able to integrate over the charge density $\rho(x_t)$
at arbitrary $F[\rho]$. The aim of this paper is to show how 
this is done.
Let us start by considering the functional integral over $\delta A$ at fixed
$\rho$ and $\rho^\prime$. Although this integration is complicated 
\cite{inprep}, the structure of the result is simple and can be understood
by simple counting of powers of the coupling constant. 
>From 
the explicit expression of $\delta\rho$ in terms of $\delta A$, 
Eqs.(\ref{rho1}),
(\ref{rho2}), it is readily seen that
\begin{eqnarray}
<\delta\rho>_{\delta A}=O(\alpha_s)\nonumber \\
<\delta\rho\delta\rho>_{\delta A}=O(\alpha_s)
\end{eqnarray}
while all other (connected) correlation functions of $\delta\rho$ are higher
order in $\alpha_s$. Since we are working to the lowest order in $\alpha_s$ we
can neglect all these other terms. It therefore follows that to lowest order,
after integrating over $\delta A$ we are left with the weight function
for $\delta \rho$, which generates only connected one- and two-point functions.
Such weight is obviously a Gaussian.
Introducing the following notations
\begin{eqnarray}
\eta & := & \alpha_s\ln(1/x)\nonumber \\
<\delta\rho^a(x_t)>_{\delta A}& =: & \eta \sigma^a(x_t)\nonumber \\
<\delta\rho^a(x_t,x^+)\delta\rho^b(y_t,x^+)>_{\delta A}& =: &
\eta \chi^{ab}(x_t,y_t)
\end{eqnarray}
we can therefore write the result of the $\delta A_\mu$ integration in the
form
\begin{eqnarray}
\lefteqn{
\int D[\rho,\rho^\prime]
[ {\rm Det}(\chi)]^{-1/2}\exp\left(-F[\rho]\right) 
} 
\nonumber \\ && \times 
\exp\left(-{1\over 2\,\eta}
  \left[\rho^\prime_x-\rho_x-
    \eta \sigma_x\right][\chi^{-1}_{x y}]\left[
    \rho^\prime_y-\rho_y-\eta\sigma_y\right]\right)
\nonumber\\ & =: & 
\int D[\rho,\rho^\prime]\exp\{-U[\rho,\rho^\prime]\}
\label{next}
\end{eqnarray}
In the above equation we adopted condensed notations: the indices $x$
stand for the set of indices and coordinates $\{x_t,a\}$, and repeated
indices are understood to be summed (integrated) over. We will use
these notations in the rest of the paper.

We note here that this result can be derived formally by introducing
 the variable $\rho^\prime$ with the help of Lagrange multiplier
\begin{equation}
\delta (\rho'-\rho-\delta \rho[\delta A]) = 
\int D[\lambda]\ e^{i\lambda (\rho'-\rho-\delta \rho[\delta A])}
\end{equation}
and subsequently integrating out $\lambda$  
in perturbation theory to order $\alpha_s$.

The calculation of $\sigma$ and $\chi$ is technically nontrivial
\cite{inprep} and is beyond the scope of this note.  Here we want to
show, that assuming those quantities are known functionals of $\rho$,
the integral over $\rho$ in Eq.(\ref{next}) can be performed explicitly.

Consider again Eq.(\ref{next}). We will assume that $F[\rho]$ is a
smooth functional of $\rho$ and that the scale of its variation is
independent of the strong coupling constant $\alpha_s$. The last
factor in Eq.(\ref{next}) however is a very steep function of $\rho$
which is sharply peaked around $\rho^\prime+O(\alpha_s)$. This is due
to the factor $1/\alpha_s$ in the Gaussian weight for $\delta\rho$. It
is therefore clear that the integral can be calculated
straightforwardly in the steepest descent approximation.  We should
also remember that we need to retain the terms up to first order in
the coupling constant.  Simple counting of the powers of $\alpha_s$
shows that to reach this accuracy one needs to take into account up to
the third order contributions in the steepest descent integration.
In other words, the expansion of the exponential factor in Eq.(\ref{next}) is
needed up to the fourth order in fluctuations around the stationary
point.

The steepest descent equation to first order in $\alpha_s$ reads
\begin{equation}
\rho^{\prime }_x-\rho_x-\eta\sigma_x=\eta
\chi_{x u}\left[{\delta F\over \delta\rho_u}+{1\over 2}{\rm
    tr}(\chi^{-1} 
{\delta\chi\over\delta\rho_u})\right ]
\label{steepd}
\end{equation}

Substituting this into Eq.(\ref{next}) we find that the logarithm of
the integrand in Eq.(\ref{next}) becomes
\begin{eqnarray}
U & = & F+{1\over 2}\rm{tr}\ln(\chi)
\\ &&
+{\eta \over 2}\left[{\delta F\over \delta\rho_u}+{1\over 2}
{\rm tr}(\chi^{-1}
{\delta\chi\over\delta\rho_u})\right ]\chi_{u v}
\left[{\delta F\over \delta\rho_v}+{1\over 2}{ \rm tr}(\chi^{-1}
{\delta\chi\over\delta\rho_v})\right ]\nonumber
\label{u}
\end{eqnarray}
In the above expression all the functionals are taken at $\rho^0$ - 
the solution of the  steepest descent equation Eq.(\ref{steepd}).

Next we have to evaluate the correction due to Gaussian fluctuations of $\rho$
around $\rho^0$. After some algebra we find
\begin{eqnarray}
{\delta^2 U\over\delta\rho_x\delta\rho_y} & = &
{1\over \eta}\chi^{-1}_{x y}+
{\delta^2 \over\delta\rho_x\delta\rho_y}\left[F+{1\over 2}
\rm{tr}\ln(\chi)\right] \\
&& + \chi^{-1}_{x u}{\delta \sigma_u\over\delta\rho_y}
+{\delta \sigma_u\over\delta\rho_x}\chi^{-1}_{u y}\nonumber\\
&& + \left[\chi^{-1}_{x u}{\delta \chi_{u v}\over\delta\rho_y}
+\chi^{-1}_{y u}{\delta \chi_{u v}\over\delta\rho_x} \right]
\left[{\delta F\over\delta\rho_v}+{1\over 2}{\rm tr}\left(\chi^{-1}
{\delta \chi\over\delta\rho_v}\right)\right]\nonumber
\label{deltau}
\end{eqnarray}
Here the argument of all the functionals is again $\rho^0$.

As noted earlier to order $\alpha_s$ we have to retain two more
corrections to the steepest descent result.  This is because the third
as well as all higher derivatives of $U$ in Eq.(\ref{next}) is of the
order $1/\alpha_s$ while the fluctuation is of order
$\alpha_s^{1/2}$.  Schematically, we can therefore write all order
$\alpha_s$ contributions as
\begin{eqnarray}
\lefteqn{
\int D[\rho] \exp\left(-U[\rho]\right)
} 
\\ & = &\exp\left(-U[\rho^0]\right)\int D[\rho]\  \exp\left(-{1\over 2}U''(\rho-\rho^0)^2-
{1\over 3!}U'''(\rho-\rho^0)^3-{1\over 4!}U''''(\rho-\rho^0)^4\right)
\nonumber\\
& = &\exp\left(-U[\rho^0]\right)\int\! D[\rho]\  
\exp \left(-{1\over 2}U''(\rho-\rho^0)^2\right)
\nonumber \\ && \qquad \times  \left[1
-{1\over 4!}U''''(\rho-\rho^0)^4
+{1\over 2}\left({1\over
    3!}U'''(\rho-\rho^0)^3\right)^2\right]\nonumber\\
\nonumber \\ & = &
\exp\left(-U[\rho_0] -{1\over 2} {\rm tr}\ln \eta U'' 
-{1\over 4!}\left(
  3\hspace{0.2cm}
  \begin{minipage}[m]{2cm} \epsfysize=0.5cm \epsfbox{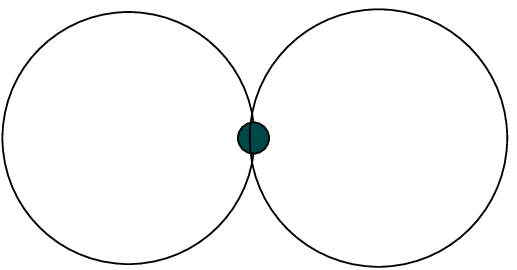}
  \end{minipage}
\right) 
+ {1\over 2 (3!)^2} \left(
  9\hspace{0.2cm}
  \begin{minipage}[m]{2cm} \epsfysize=0.5cm \epsfbox{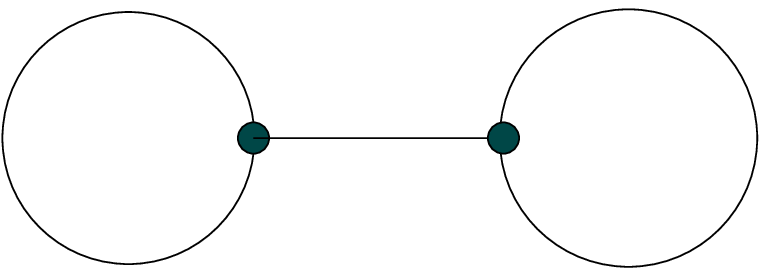}
  \end{minipage}
 \hspace{0.2cm}
  + 6 \hspace{0.2cm}   \begin{minipage}[m]{2cm}
    \epsfysize=0.5cm \epsfbox{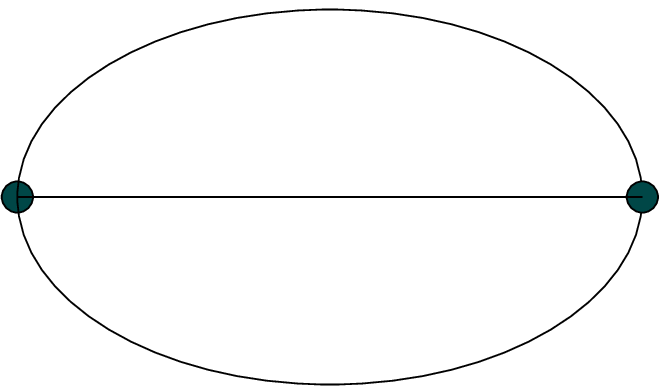} \end{minipage}
 \right)\right)
\end{eqnarray}
Here we have used the obvious simplified notations and graphical
representation in terms of the ``propagator'' $[U'']^{-1}$ and three -  and
four - point vertices $U'''$ and $U''''$.
We have also re-exponentiated the result of the
integration which is valid to leading order in $\alpha_s$.

Our task is simplified, however by the fact that we need to retain
only the leading contribution to this term. That for $U'''$ and
$U''''$ we only need to keep the pieces of order $1/\alpha_s$, and for
$<(\rho-\rho^0)^n>$ only pieces of order $\alpha_s^{n/2}$.  Some
straightforward algebra yields
\begin{eqnarray}
\label{four}
F_4 & := & {3\over 4!}
  \hspace{0.2cm}
  \begin{minipage}[m]{2cm} \epsfysize=0.5cm \epsfbox{eight.eps}
  \end{minipage}
=
{3\over 4!}{\delta^4 U\over \delta\rho_u\delta\rho_v
\delta\rho_w\delta\rho_z}
\left[{\delta^2U\over \delta\rho\delta\rho}\right]^{-1}_{u v}
\left[{\delta^2U\over \delta\rho\delta\rho}\right]^{-1}_{w z}\\
& = & \eta\left[-{1\over 4}{\rm tr}(\chi^{-1}
{\delta^2\chi\over\delta\rho_u\delta\rho_v})\chi_{u v}
+{1\over 2}{\rm tr}
(\chi^{-1}{\delta\chi\over\delta\rho_u}
\chi^{-1}{\delta\chi\over\delta\rho_v})\chi_{u v}
\right. \nonumber\\  && \hspace{1cm} \left.
-{1\over 2}{\delta^2\chi_{u v}\over\delta\rho_u\delta\rho_v}
+{1\over 2}{\delta\chi_{u w}\over\delta\rho_u}\chi^{-1}_{w
  z}{\delta\chi_{z v}
\over\delta\rho_v}
+{1\over 2}{\delta\chi_{u w}\over\delta\rho_v}\chi^{-1}_{w
  z}{\delta\chi_{z v}
\over\delta\rho_u}\right]\nonumber
\end{eqnarray}

and
\begin{eqnarray}
\label{three}
F_3 & := &
-{1\over 2 (3!)^2} \left(
  9\hspace{0.2cm}
  \begin{minipage}[m]{2cm} \epsfysize=0.5cm \epsfbox{weights.eps}
  \end{minipage}
 \hspace{0.2cm}
  + 6 \hspace{0.2cm}   \begin{minipage}[m]{2cm}
    \epsfysize=0.5cm \epsfbox{oval.eps} \end{minipage}
 \right)
\\ & = &
-{1\over 2}({1\over 3!})^2
{\delta^3 U\over \delta\rho_u\delta\rho_v\delta\rho_w}
{\delta^3 U\over \delta\rho_x\delta\rho_y\delta\rho_z}
\nonumber  \\ &&
\times\left(9\left[{\delta^2U\over \delta\rho\delta\rho}\right]^{-1}_{u v}
\left[{\delta^2U\over \delta\rho\delta\rho}\right]^{-1}_{w x}
\left[{\delta^2U\over \delta\rho\delta\rho}\right]^{-1}_{y z}
\right. \nonumber\\ && \left. \hspace{1cm}
+6\left[{\delta^2U\over \delta\rho\delta\rho}\right]^{-1}_{u x}
\left[{\delta^2U\over \delta\rho\delta\rho}\right]^{-1}_{v y}
\left[{\delta^2U\over \delta\rho\delta\rho}\right]^{-1}_{w z}\right)
\nonumber \\
&&=-\eta\left[{1\over 8}{\rm tr}
(\chi^{-1}{\delta\chi\over\delta\rho_u})\chi_{u v}{\rm tr}
(\chi^{-1}{\delta\chi\over\delta\rho_v})
+{1\over 4}{\rm tr}
(\chi^{-1}{\delta\chi\over\delta\rho_u}\chi^{-1}
{\delta\chi\over\delta\rho_v})\chi_{u v}\}
\right. \nonumber\\ &&\left. \hspace{1cm} 
+{1\over 2}{\rm tr}
(\chi^{-1}{\delta\chi\over\delta\rho_u})
{\delta\chi_{u v}\over\delta\rho_v}
+{1\over 2}{\delta\chi_{u v}\over\delta\rho_u}\chi^{-1}_{v w}
{\delta\chi\over\delta\rho_z}
+{1\over 2}{\delta\chi_{u v}\over\delta\rho_z}\chi^{-1}_{v w}
{\delta\chi\over\delta\rho_u}\right]
\nonumber 
\end{eqnarray}

Putting Eqs.(\ref{u}), (\ref{deltau}), (\ref{four}) and (\ref{three})
together and eliminating the stationary value of $\rho$ in
favor of $\rho^\prime$ through Eq.(\ref{steepd}), we find
\begin{eqnarray}
\label{finals}
&&F^\prime=U+{1\over 2}{\rm tr}\ln {\delta^2
  U\over\delta\rho_x\delta\rho_y }+F_3+F_4\\ 
&&=F+{\alpha_s\ln(1/x)\over 2}\left[\chi_{u v}
{\delta^2 \over\delta\rho_u\delta\rho_v}F
-{\delta^2\chi_{u v} \over\delta\rho_u\delta\rho_v}
-{\delta F\over \delta\rho_u}
\chi_{u v}
{\delta F\over \delta\rho_v}
\right. \nonumber \\ && \left. \hspace{1cm}
+2{\delta F\over \delta\rho_u}{\delta\chi_{u v}\over
\delta\rho_v}+2{\delta \sigma_u\over\delta\rho_u}
-2{\delta F\over\delta\rho_u}\sigma_u\right]\nonumber
\end{eqnarray}

Equation (\ref{finals}) gives the Wilson renormalization group equation.
\begin{eqnarray}
\label{finalrg}
{d\over d\ln(1/x)}F
& = &{\alpha_s\over 2}\left[\chi_{u v}
{\delta^2 \over\delta\rho_u\delta\rho_v}F
-{\delta^2\chi_{u v} \over\delta\rho_u\delta\rho_v}
-{\delta F\over \delta\rho_u}
\chi_{u v}
{\delta F\over \delta\rho_v}
\right. \nonumber\\ && \left. \hspace{1cm}
+
2{\delta F\over \delta\rho_u}{\delta\chi_{u v}\over
\delta\rho_v}
+2{\delta \sigma_u\over\delta\rho_u}
-2{\delta F\over\delta\rho_u}\sigma_u\right]
\end{eqnarray}
This equation is extremely simple when written for the weight function
$Z\equiv \exp\{-F\}$
\begin{equation}
{d\over d\ln(1/x)}Z=
\alpha_s
\left\{{1\over 2}{\delta^2
    \over\delta\rho_u\delta\rho_v}\left[Z\chi_{u v}
\right]
-{\delta \over\delta\rho_u}\left[Z\sigma_u\right]\right\}
\label{final}
\end{equation}

Equations (\ref{finalrg}) and (\ref{final}) are the central result of
this paper.  They provide the closed form of the renormalization group
equation in terms of the functionals $\sigma[\rho]$ and $\chi[\rho]$.

We want to make now several remarks. 
First, the two terms in Eqs.(\ref{final}) have a natural interpretation
as the real and the virtual contributions to the evolution. The average
fluctuation $\chi$ is generated by the real ladder - like diagrams
of the gluon fluctuation $\delta A$ while the average color density
comes from the virtual one loop diagrams \cite{inprep}.

Second, Eq.(\ref{final}) can be
written directly as evolution equation for the correlators of the
charge density. Multiplying Eq.(\ref{final}) by
$\rho_{x_1}...\rho_{x_n}$ and integrating over $\rho$ yields
\begin{eqnarray}
\lefteqn{
{d\over d\ln(1/x)}<\rho_{x_1}...\rho_{x_n}>
}
\\ &&=\alpha_s
\left[\sum_{0<m<k<n+1}<\rho_{x_1}...\rho_{x_{m-1}}\rho_{x_{m+1}}... 
\rho_{x_{k-1}}\rho_{x_{k+1}}...
\rho_{x_n}\chi_{x_m x_k}>
\right. \nonumber\\ && \left. \hspace{1cm}
+\sum_{0<l<n+1}<\rho_{x_1}...\rho_{x_{l-1}}\rho_{x_{l+1}}
...\rho_{x_n}\sigma_{x_l}>\right]
\nonumber 
\label{correl}
\end{eqnarray}

In particular, taking $n=2$ we obtain the evolution equation for the two
point function
\begin{equation}
{d\over d\ln(1/x)}<\rho_x\rho_y>=
\alpha_s\left\{<\chi_{x y}+\rho_x\sigma_y+\rho_y\sigma_x>\right\}
\label{prop}
\end{equation}
In general this is not a closed equation since $\sigma$ and $\chi$ depend
on $\rho$ in a complicated nonlinear way. However in the weak field limit,
$\sigma$  and $\chi$ are linear and quadratic functionals of $\rho$ 
respectively.
In this case Eq.(\ref{prop}) becomes a closed 
equation for the two point correlation
function of $\rho$. The explicit expressions for $\sigma$ and $\chi$ in this
limit have been calculated in \cite{linearRG} where it is also shown that
Eq.(\ref{prop}) is precisely the
famous BFKL equation.

 Since the correlator of the colour charge density 
is directly related to the unintegrated gluon density
in a hadron (nucleus) \cite{linearRG}, Eq. (\ref{prop}) can be 
straightforwardly rewritten as a
nonlinear evolution equation for the gluon
density. It would be very interesting to compare this equation
to the generalized evolution equation derived by Levin and Laenen \cite{ll}. 
In the second order in the gluon density (and neglecting the
virtual contributions in the double logarithmic approximation)
Eq. (\ref{prop}) should reduce to the famous GLR
equation \cite{GLR} with the coefficient calculated by
Mueller and Qiu \cite{mq}.

Another remark we want to make is that a general
Gaussian form of the weight function
\begin{equation}
F=\int dx_tdy_t \rho(x_t)\mu^{-1}(x_t,y_t)\rho(y_t)
\label{gauss}
\end{equation} 
is not an eigenfunction
of equation (\ref{finalrg}). This form of $F$ was argued in \cite{larry} to
be a good approximation for valence partons in a large enough
nucleus, due to incoherence of color charges of partons coming from different
nucleons. In general it should also be a good approximation in perturbative
regime where $F$ can be expanded in powers of $\rho$. We see however, 
that at low enough $x$ non quadratic terms are generated through the evolution
Eq.(\ref{finalrg}). In particular, even if we take the weak field expressions
for $\sigma$ and $\chi$, the term quadratic in $F$ on the r.h.s. of 
Eq.(\ref{finalrg}) generates a correction quartic in $\rho$. Obviously, at
weak coupling and small $\rho$ this term can be neglected (or rather
approximated by a quadratic term in a mean field like manner) so that a 
Gaussian $Z$ is recovered in the perturbative regime. 
This suggests that a study of the RG flow Eq.(\ref{finalrg}) should indeed
use Eq.(\ref{gauss}) as an initial condition for the evolution starting 
at not too small values of $x$. The natural choice for the initial
value of
$\mu(x_t,y_t)$ is 
\begin{equation}
\mu(x_t,y_t)=S(b_t)\int {d^2k_t\over (2\pi)^2}\ 
e^{ik_t(x_t-y_t)}\, \phi(k_t,x_0)
\end{equation}
Here $b_t={x_t+y_t\over 2}$ is the impact parameter, $S(b_t)$ is a
nucleon shape function, $x_0$ is the value of $x$ from which we start evolving
$F$ according to Eq.{\ref{finalrg} and $\phi(k_t,x)$ is the unintegrated 
gluon density.

Finally, we note that $\sigma$ and $\chi$ are indeed calculable in a closed
form. The result of this calculation will be presented elsewhere \cite{inprep}.

{\bf Acknowledgements} We are grateful to Larry McLerran for numerous
discussions on a variety of topics related to the subject of this
paper.  This work was started during the program ``Ultrarelativistic
Nuclei: from structure functions to the quark-gluon plasma'' at INT,
University of Washington. We thank INT for hospitality and financial
support and the participants of the program for enjoyable atmosphere
and very lively and stimulating discussions.  The work of J.J.M. and
A.K. was supported by DOE contracts DOE-Nuclear DE-FG02-87ER-40328 and
DOE High Energy DE-AC02-83ER40105 respectively.  The work of A.L. was
partially supported by Russian Fund for Basic Research, Grant
96-02-16210. HW was supported by the EC Programme ``Training and
Mobility of Researchers", Network ``Hadronic Physics with High Energy
Electromagnetic Probes", contract ERB FMRX-CT96-0008.

\end{document}